\documentclass[11pt]{article}
\usepackage{graphicx}
\usepackage{amsmath}
\usepackage{amsfonts}
\usepackage{amssymb}

\newcommand{\be}{\begin{equation}}
\newcommand{\ee}{\end{equation}}
\newcommand{\bq}{\begin{eqnarray}}
\newcommand{\eq}{\end{eqnarray}}

\begin{document}

\title{\textbf{The Bohm Interpretation of Quantum Cosmology}}
\author{\ \textbf{\ \ Nelson Pinto-Neto }\\{\small \ \ \ }\\
{\small \ \ Centro Brasileiro de Pesquisas F\'{i}sicas, CBPF \ \ \ }\\
{\small Rua Dr. Xavier Sigaud 150 , 22290-180, Rio de Janeiro,
Brazil; e-mail: nelsonpn@cbpf.br \ \ \ \ \ \ \ \ \ \ \ }}
\maketitle

I make a review on the aplications of the Bohm-De Broglie interpretation
of quantum mechanics to quantum cosmology. In the framework of minisuperspaces
models, I show how quantum cosmological
effects in Bohm's view can avoid the initial singularity, isotropize
the Universe, and even be a cause for the present observed acceleration of
the Universe. In the general case, we enumerate the possible structures
of quantum space and time.

\section{INTRODUCTION}

It is a great honour for me to write
this contribution in memory of Prof. James T. Cushing.
I met Prof. Cushing during a symposium in S\~ao Paulo
about Bohm's theory in 1999, where I could appreciate the clarity of his ideas, and
his ability to find out the main point in a controversy.
A gentleman and a deep thinker, capable to put in clear terms any difficult
subject on his domain of interest.

One of the main topics of research of Prof. Cushing was the Bohm-De Broglie
interpretation of quantum mechanics in its many aspects, and its
comparison with other formulations of quantum mechanics. At the moment,
there is no clear observation selecting one of these formulations.
Hence, it is desirable to push these formulations to the frontiers
of physics in order to discriminate them either by a possible
feasible experiment or by a matter of principle (self consistency,
resolution of fundamental issues, etc...).

One of these frontiers is
cosmology, a unique domain in physics, where the totality of all physical
systems, including spacetime itself and its geometry, is under investigation,
composing a single system which cannot be manipulated or prepared, it can
only be observed, and containing the observers themselves. The proposal
of this contribution is to review the aplications of the Bohm-De Broglie
interpretation to quantum cosmology, see if it helps
in the resolution of the traditional cosmological issues, and
compare their results with the ones
obtained in other formulations of quantum mechanics.

Two of the main questions one might ask in cosmology are:

1) Is the Universe eternal or it had a beginning, and in the last case, was
this beginning given by an initial singularity?

2) Why the Universe we live in is remarkably homogeneous and isotropic, with
very small deviations from this highly symmetric state?

The answer given by classical General Relativity (GR) to the first question,
indicated by the singularity theorems \cite{he}, asserts that probably the
Universe had a singular beginning. As singularities are out of the scope of
any physical theory, this answer invalidates any description of the very
beginning of the Universe in physical terms. One might think that GR and/or
any other matter field theory must be changed under the extreme situations
of very high energy density and curvature near the singularity, rendering
the physical assumptions of the singularity theorems invalid near this
point. One good point of view (which is not the only one) is to think that
quantum gravitational effects become important under these extreme
conditions. We should then construct a quantum theory of gravitation and
apply it to cosmology. For instance, in the euclidean quantum gravity
approach \cite{haw} to quantum cosmology, a second answer to the first
question comes out: the Universe may have had a non-singular birth given by
the beginning of time through a change of signature.

In the same way, the naive answer of GR and the standard cosmological
scenario to the second question is not at all satisfactory: the reason for
the Universe be highly homogeneous and isotropic is a matter of initial
conditions. However, solutions of Einstein's equations with this symmetry
are of measure zero; so, why the Universe is not inhomogeneous and/or
anisotropic? Inflation \cite{gut,kol} is an idea that tries to explain this
fact. Nevertheless, in order for inflation to happen some special initial
conditions are still necessary, although much more less stringent than in
the case without inflation. Once again, quantum cosmology can help in this
matter by providing the physical reasons for having the initial conditions
for inflation.

Does it makes sense to quantize the whole Universe?
Almost all physicists believe that quantum mechanics is a universal and
fundamental theory, applicable to any physical system, from which
classical physics can be recovered.  The Universe is, of course, a
valid physical system:  there is a theory, Standard Cosmology, which is
able to describe it in physical terms, and make predictions which can
be confirmed or refuted by observations. In fact, the observations
until now confirm the standard cosmological scenario with a cosmological
constant. Hence, supposing
the universality of quantum mechanics, the Universe itself must be
described by quantum theory, from which we could recover Standard
Cosmology. However, the Copenhaguen interpretation of quantum mechanics
\cite{bohr,hei,von}, which is the one
taught in undergraduate courses and employed by the majority of
physicists in all areas (specially the von Neumann's approach), cannot
be used in a Quantum Theory of Cosmology. This is because it imposes
the existence of a classical domain. In von Neumann's view, for
instance, the necessity of a classical domain comes from the way it
solves the measurement problem (see Ref. \cite{omn} for a good
discussion).  In an impulsive measurement of some observable, the wave
function of the observed system plus macroscopic apparatus splits into
many branches which almost do not overlap (in order to be a good
measurement), each one containing the observed system in an eigenstate
of the measured observable, and the pointer of the apparatus pointing
to the respective eigenvalue. However, in the end of the measurement,
we observe only one of these eigenvalues, and the measurement is robust
in the sense that if we repeat it immediately after, we obtain the same
result. So it seems that the wave function collapses, the other
branches disappear. The Copenhaguen interpretation assumes that this
collapse is real.  However, a real collapse cannot be described by the
unitary Schr\"{o}dinger  evolution. Hence, the Copenhaguen
interpretation must assume that there is a fundamental process in a
measurement which must occur outside the quantum world, in a classical
domain.  Of course, if we want to quantize the whole Universe, there is
no place for a classical domain outside it, and the Copenhaguen
interpretation cannot be applied.  Hence, if someone insists with the
Copenhaguen interpretation, she or he must assume that quantum theory
is not universal, or at least try to improve it by means of further
concepts. One possibility is by invoking the
phenomenon of decoherence \cite{deco}. In fact, the interaction of
the observed quantum
system with its environment yields an effective diagonalization of the
reduced density matrix, obtained by tracing out the irrelevant degrees of
freedom. Decoherence can
explain why the splitting of the wave function is given in terms of the
pointer basis states, and why we do not see superpositions of
macroscopic objects. In this way, classical properties emerge from
quantum theory without the need of being assumed.  In the framework of
quantum gravity, it can also explain how a classical background
geometry can emerge in a quantum universe \cite{2kie}. In fact, it is the first
quantity to become classical. However, decoherence is not yet a
complete answer to the measurement problem \cite{muk,zeh}. It does not
explain the apparent collapse after the measurement is completed, or
why all but one of the diagonal elements of the density matrix become
null when the measurement is finished. The theory is unable to give an
account of the existence of facts, their uniqueness as opposed to the
multiplicity of possible phenomena. Further developments are still in progress, like the consistent
histories approach \cite{har}, which is however incomplete until now.
The important role
played by the observers in these descriptions is not yet explained \cite{zur},
and still remains the problem on how to describe a quantum universe
when the bakground geometry is not yet classical.

Nevertheless, there are
some alternative solutions to this quantum cosmological dilemma which, together with decoherence,
can solve the measurement problem maintaining the universality of quantum
theory. One can say that the Schr\"{o}dinger
evolution is an approximation of a more fundamental non-linear
theory which can accomplish the collapse \cite{rim,pen}, or that the
collapse is effective but not real, in the sense that the other
branches disappear from the observer but do not disappear from
existence. In this second category we can cite the Many-Worlds
Interpretation \cite{eve} and the Bohm-de Broglie Interpretation
\cite{bohm,hol,vig,val2}. In the former, all the possibilities in the splitting
are actually realized. In each branch there is an observer with the
knowledge of the corresponding eigenvalue of this branch, but she or he
is not aware of the other observers and the other possibilities because
the branches do not interfere. In the latter, a point-particle in
configuration space describing the observed system and apparatus is
supposed to exist, independently on any observations.  In the
splitting, this point particle will enter into one of the branches
(which one depends on the initial position of the point particle before
the measurement, which is unknown), and the other branches will be
empty. It can be shown \cite{hol} that the empty waves can neither
interact with other particles, nor with the point particle containing
the apparatus.  Hence, no observer can be aware of the other branches
which are empty.  Again we have an effective but not real collapse (the
empty waves continue to exist), but now with no multiplication of
observers. Of course these interpretations can be used in quantum
cosmology. Schr\"{o}dinger  evolution is always valid, and there is no need
of a classical domain outside the observed system.

In this contribution we will foccus on the application of the Bohm-de Broglie
interpretation to quantum cosmology \cite{vink,sht,val,bola1}. In this
approach, the fundamental object of quantum gravity, the geometry of
3-dimensional spacelike hypersurfaces, is supposed to exist independently on
any observation or measurement, as well as its canonical momentum, the extrinsic
curvature of the spacelike hypersurfaces. Its evolution, labeled by some time
parameter, is dictated by a quantum evolution that is different from
the classical one due to the presence of a quantum potential which
appears naturally from the Wheeler-DeWitt equation. This interpretation
has been applied to many minisuperspace models
\cite{vink,bola1,kow,hor,bola2,fab,fab2,iso}, obtained by the imposition of
homogeneity of the spacelike hypersurfaces.  The classical limit, the
singularity problem, the cosmological constant problem,
and the time issue have been discussed. For instance, in some of these
papers it was shown that in models involving scalar fields or
radiation, which are good representatives of the matter content of the
early universe, the singularity can be clearly avoided by quantum
effects. In the Bohm-de Broglie interpretation description, the quantum
potential becomes important near the singularity, yielding a repulsive
quantum force counteracting the gravitational field, avoiding the
singularity and yielding inflation. The classical limit (given by the
limit where the quantum potential becomes negligible with respect to
the classical energy) for large scale factors are usually
attainable, but for some scalar field models it depends on the quantum
state and initial conditions (see Ref.\cite{hartle}).
In fact it is possible to have small
classical universes and large quantum ones \cite{fab}.
For instance, in Ref.\cite{fab2} we obtained models arising classically from
a singularity, experiencing quantum effects in the middle of their
expansion, and recovering their classical behaviour for large values of
$\alpha$. These quantum effects may cause an acceleration of the expansion
of these models as the one which is observed in our Universe. Such a possibility
is explored in detail in Ref.\cite{acc} and shown to be a viable alternative
explanation for th recent high redshift supernvae observations \cite{SN1,SN}
About the time
issue (different choices of time yielding different physical predictions
\cite{isham,lemos}), it was shown that for any choice of the lapse function the
quantum evolution of the homogeneous hypersurfaces yield the same
four-geometry \cite{bola1}. In Ref.\cite{iso} we have shown how
quantum cosmological effects in the Bohmian description may not only
avoid the initial singularity but also can isotropize anisotropic models
which classically never isotropizes.

Finally, we studied the bohmian view of quantum space and time in the full theory, where we are not restricted to
homogeneous spacelike hypersurfaces \cite{euc}.  The question is, given an initial
hypersurface with consistent initial conditions, does the evolution of
the initial three-geometry driven by the quantum bohmian dynamics yields
the same four-geometry for any choice of the lapse and shift functions, and
if it does, what kind of spacetime structure is formed? We
know that this is true if the three-geometry is evolved by the dynamics of
classical General Relativity (GR), yielding a non degenerate four geometry,
but it can be false if the evolving
dynamics is the quantum bohmian one. Our conclusion is that, in general, the quantum bohmian
evolution of the three-geometries does not yield any non degenerate
four-geometry at all.
Only for very special quantum states a relevant quantum non degenerate
four-geometry can be
obtained, and it must be euclidean. In the general case, either the quantum bohmian
evolution is consistent (still independent on the choice
of the lapse and shift functions) but yielding a degenerate
four-geometry, where special vector fields, the null eigenvectors of the
four geometry, are present\footnote{For instance, the four geometry of
Newtonian spacetime is degenerate \cite{new}, and its single null eigenvector
is the normal
of the absolute hypersurfaces of simultaneity, the time. As we know,
it does not
form a single spacetime structure because it is broken in absolute space plus
absolute time.}. We arrive at these
conclusions without assuming any regularization and
factor ordering of the Wheeler-DeWitt equation. As we know, the
Wheeler-DeWitt equation involves the application of the product of
local operators on states at the same space point, which is ill defined
\cite{reg}.  Hence we need to regularize it in order to
solve the factor ordering problem, and have a theory free of anomalies
(for some proposals, see Refs \cite{japa1,japa2,kow2}). Our conclusions
are completly independent on these issues. Also, in the general
case where there are degenerate
four-geometries, we can obtain a picture of the
quantum structure yielded by the bohmian dynamics, which is not a
spacetime in the sense described above but something else, as the degenerate four-geometries compatible
with the Carroll group \cite{poin}.

This contribution is organized as follows: in the next section
I present the Bohm-De Broglie
interpretation of canonical quantum gravity and its restriction to the
minisuperspace of homogeneous geometries. I show that, in this framework,
the time issue can be avoided in the bohmian approach to quantum cosmology.
In section III I study the singularity and isotropy problems in minisuperspace
quantum cosmology. In section IV I
treat the full superspace and the quantum picture of space
and time coming from Bohm's point of view. We end up with the conclusions.

\section{THE BOHM-DE BROGLIE INTERPRETATION OF CANONICAL QUANTUM COSMOLOGY}

Let me now apply the Bohm-de Broglie interpretation to canonical quantum
cosmology. I will
quantize General Relativity Theory (GR) where the matter content is constituted of a
minimally coupled scalar field with arbitrary potential. All subsequent
results remain essentially the same for any matter field which couples
uniquely with the metric, not with their derivatives.

The classical hamiltonian of GR with a scalar field is given by:
\begin{equation}
\label{hgr}
H = \int d^3x(N{\cal H}+N^j{\cal H}_j)
\end{equation}
where
\begin{eqnarray}
\label{h0}
{\cal H} &=& \kappa G_{ijkl}\Pi ^{ij}\Pi ^{kl} +
\frac{1}{2}h^{-1/2}\Pi ^2 _{\phi}+\nonumber\\
& & + h^{1/2}\biggr[-{\kappa}^{-1}(R^{(3)} - 2\Lambda)+
\frac{1}{2}h^{ij}\partial _i \phi\partial _j \phi+U(\phi)\biggl]\\
\label{hi}
{\cal H}_j &=& -2D_i\Pi ^i_j + \Pi _{\phi} \partial _j \phi .
\end{eqnarray}
In these equations, $h_{ij}$ is the metric of closed 3-dimensional
spacelike hypersurfaces, and $\Pi ^{ij}$ is its canonical momentum
given by
\begin{equation}
\label{ph}
\Pi ^{ij} = - h^{1/2}(K^{ij}-h^{ij}K) =
G^{ijkl}({\dot{h}}_{kl} -  D _k N_l - D _l N_k ),
\end{equation}
where
\begin{equation}
K_{ij} = -\frac{1}{2N} ({\dot{h}}_{ij} -  D _i N_j - D _j N_i ) ,
\end{equation}
is the extrinsic curvature of the hypersurfaces (indices are raisen and lowered
by the 3-metric $h_{ij}$ and its inverse $h^{ij}$). The canonical momentum
of the scalar field is
\begin{equation}
\label{pf}
\Pi _{\phi} = \frac{h^{1/2}}{N}\biggr(\dot{\phi}-N^i \partial _i \phi \biggl).
\end{equation}
The quantity $R^{(3)}$ is the
intrinsic curvature of the hypersurfaces and $h$ is the determinant
of $h_{ij}$.
The lapse function $N$ and the shift function $N_j$ are the
Lagrange multipliers of the super-hamiltonian constraint
${\cal H}\approx 0$ and the super-momentum constraint
${\cal H}^j \approx 0$,
respectively. They are present due to the invariance of GR under
spacetime coordinate transformations. The quantities $G_{ijkl}$ and
its inverse $G^{ijkl}$ ($G_{ijkl}G^{ijab}=\delta ^{ab}_{kl}$) are
given by
\begin{equation}
\label{300}
G^{ijkl}=\frac{1}{2}h^{1/2}(h^{ik}h^{jl}+h^{il}h^{jk}-2h^{ij}h^{kl}),
\end{equation}
\begin{equation}
\label{301}
G_{ijkl}=\frac{1}{2}h^{-1/2}(h_{ik}h_{jl}+h_{il}h_{jk}-h_{ij}h_{kl}),
\end{equation}
which is called the DeWitt metric. The quantity $D_i$ is the $i$-component
of the covariant derivative operator on the hypersurface, and
$\kappa = 16 \pi G/c^4$.

The classical 4-metric
\begin{equation}
\label{4g}
ds^{2}=-(N^{2}-N^{i}N_{i})dt^{2}+2N_{i}dx^{i}dt+h_{ij}dx^{i}dx^{j}
\end{equation}
and the scalar field  which are solutions
of the Einstein's equations can be obtained from the Hamilton's equations
of motion
\begin{equation}
\label{hh}
{\dot{h}}_{ij} = \{h_{ij},H\},
\end{equation}
\begin{equation}
\label{hp}
{\dot{\Pi}}^{ij} = \{\Pi ^{ij},H\},
\end{equation}
\begin{equation}
\label{hf}
{\dot{\phi}} = \{\phi,H\},
\end{equation}
\begin{equation}
\label{hpf}
{\dot{\Pi _{\phi}}}= \{\Pi _{\phi},H\},
\end{equation}
for some choice of $N$ and $N^i$, and
if we impose initial conditions compatible with the constraints
\begin{equation}
\label{hh0}
{\cal H} \approx 0 ,
\end{equation}
\begin{equation}
\label{hhi}
{\cal H}_i \approx 0.
\end{equation}
It is a feature of the hamiltonian of GR that the 4-metrics (\ref{4g})
constructed in this way, with the same initial conditions, describe the
same four-geometry for any choice of $N$ and $N^i$.

The algebra of the constraints close in the following form
(we follow the notation of Ref. \cite{kuc1}):

\begin{eqnarray}\label{algebra}
\{ {\cal H} (x), {\cal H} (x')\}&=&{\cal H}^i(x) {\partial}_i \delta^3(x,x')-
{\cal H}^i(x'){\partial}_i \delta^3(x',x) \nonumber \\
\{{\cal H}_i(x),{\cal H}(x')\}&=&{\cal H}(x) {\partial}_i \delta^3(x,x')  \\
\{{\cal H}_i(x),{\cal H}_j(x')\}&=&{\cal H}_i(x) {\partial}_j \delta^3(x,x')+
{\cal
H}_j(x'){\partial}_i \delta^3(x,x') \nonumber
\end{eqnarray}

To quantize this constrained system, we follow the Dirac quantization
procedure. The constraints
become conditions imposed on the possible states of the quantum
system, yielding the following quantum equations:
\begin{eqnarray}
\label{smo}
\hat{{\cal H}}_i \mid \Psi  \! > &=& 0 \\
\label{wdw0}
\hat{{\cal H}} \mid \Psi  \! > &=& 0
\end{eqnarray}
In the metric and field representation, the first equation is
\begin{equation}
\label{smo2}
-2 h_{li}D_j\frac{\delta \Psi(h_{ij},\phi)}{\delta h_{lj}} +
\frac{\delta \Psi(h_{ij},\phi)}{\delta \phi} \partial _i \phi = 0 ,
\end{equation}
which implies that the wave functional $\Psi$ is an invariant under
space coordinate transformations.

The second equation is the Wheeler-DeWitt equation \cite{whe,dew}. Writing
it unregulated in the coordinate representation we get
\begin{equation}
\label{wdw2}
\biggr\{-\hbar ^2\biggr[\kappa G_{ijkl}\frac{\delta}{\delta h_{ij}}
\frac{\delta}{\delta h_{kl}}
 + \frac{1}{2}h^{-1/2} \frac{\delta ^2}{\delta \phi ^2}\biggl] +
V\biggl\}\Psi(h_{ij},\phi) = 0 ,
\end{equation}
where $V$ is the classical potential given by
\begin{equation}
\label{v}
V = h^{1/2}\biggr[-{\kappa}^{-1}(R^{(3)} - 2\Lambda)+
\frac{1}{2}h^{ij}\partial _i \phi\partial _j \phi+
U(\phi)\biggl] .
\end{equation}
This equation involves products of local operators at the same
space point, hence it must be regularized. After doing this, one
should find a factor ordering which makes the theory free of anomalies,
in the sense that the commutator of the operator version of the constraints
close in the same way as their respective classical Poisson brackets
(\ref{algebra}). Hence, Eq. (\ref{wdw2}) is only a formal one which must be
worked out \cite{japa1,japa2,kow2}.

Let us now see what is the Bohm-de Broglie interpretation of the solutions of
Eqs. (\ref{smo2}) and (\ref{wdw2}) in the metric and field representation.
First we write the wave functional in polar form
$\Psi = A\exp (iS/\hbar )$, where $A$ and $S$ are functionals of
$h_{ij}$ and $\phi$. Substituting it in Eq. (\ref{smo2}), we get two
equations saying that $A$ and $S$ are invariant under general space
coordinate transformations:
\begin{equation}
\label{smos}
-2 h_{li}D_j\frac{\delta S(h_{ij},\phi)}{\delta h_{lj}} +
\frac{\delta S(h_{ij},\phi)}{\delta \phi} \partial _i \phi = 0 ,
\end{equation}
\begin{equation}
\label{smoa}
-2 h_{li}D_j\frac{\delta A(h_{ij},\phi)}{\delta h_{lj}} +
\frac{\delta A(h_{ij},\phi)}{\delta \phi} \partial _i \phi = 0 .
\end{equation}

The two equations we obtain for $A$ and $S$ when we substitute
$\Psi = A\exp (iS/\hbar )$ into Eq. (\ref{wdw2}) will of course
depend on the factor ordering we choose. However, in any case,
one of the equations will have the form
\begin{equation}
\label{hj}
\kappa G_{ijkl}\frac{\delta S}{\delta h_{ij}}
\frac{\delta S}{\delta h_{kl}}
 + \frac{1}{2}h^{-1/2} \biggr(\frac{\delta S}{\delta \phi}\biggl)^2
+V+Q=0 ,
\end{equation}
where $V$ is the classical potential given in Eq. (\ref{v}).
Contrary to the other terms in Eq. (\ref{hj}),
which are already well defined, the precise form of $Q$ depends on the regularization
and factor ordering which are prescribed for the Wheeler-DeWitt equation.
In the unregulated form given in Eq. (\ref{wdw2}), $Q$ is
\begin{equation}
\label{qp1}
Q = -{\hbar ^2}\frac{1}{A}\biggr(\kappa G_{ijkl}\frac{\delta ^2 A}
{\delta h_{ij} \delta h_{kl}} + \frac{h^{-1/2}}{2} \frac{\delta ^2 A}
{\delta \phi ^2}\biggl) .
\end{equation}
Also, the other equation besides (\ref{hj}) in this case is
\begin{equation}
\label{pr}
\kappa G_{ijkl}\frac{\delta}{\delta h_{ij}}\biggr(A^2
\frac{\delta S}{\delta h_{kl}}\biggl)+\frac{1}{2}h^{-1/2}
\frac{\delta}{\delta \phi}\biggr(A^2
\frac{\delta S}{\delta \phi}\biggl) = 0 .
\end{equation}

Let me now implement the Bohm-de Broglie interpretation for canonical quantum
gravity.  First of all we note that Eqs. (\ref{smos}) and (\ref{hj}),
which are always valid irrespective of any factor ordering of the
Wheeler-DeWitt equation, are like the Hamilton-Jacobi equations for GR,
suplemented by an extra term $Q$ in the case of Eq. (\ref{hj}), which
we will call the quantum potential. By analogy with the cases of
non-relativistic particle and quantum field theory in flat spacetime, we will
postulate that the 3-metric of spacelike hypersurfaces, the scalar
field, and their canonical momenta always exist, independent on any
observation, and that the evolution of the 3-metric and scalar field
can be obtained from the guidance relations
\begin{equation}
\label{grh}
\Pi ^{ij} = \frac{\delta S(h_{ab},\phi)}{\delta h_{ij}} ,
\end{equation}
\begin{equation}
\label{grf}
\Pi _{\phi} = \frac{\delta S(h_{ij},\phi)}{\delta \phi} ,
\end{equation}
with $\Pi ^{ij}$ and $\Pi _{\phi}$ given by Eqs. (\ref{ph}) and
(\ref{pf}), respectively. Like before, these are first order
differential equations which can be integrated to yield the 3-metric
and scalar field for all values of the $t$ parameter. These solutions
depend on the initial values of the 3-metric and scalar field at some
initial hypersurface.  The evolution of these fields will of course be
different from the classical one due to the presence of the quantum
potential term $Q$ in Eq. (\ref{hj}).  The classical limit is once more
conceptually very simple: it is given by the limit where the quantum
potential $Q$ becomes negligible with respect to the classical energy.
The only difference from the cases of the non-relativistic
particle and quantum field theory in flat spacetime is the fact that
Eq. (\ref{pr}) for canonical quantum gravity cannot
be interpreted as a continuity equation for a probabiblity density $A^2$
because of the hyperbolic nature of the DeWitt metric $G_{ijkl}$.
However, even without a notion of probability, which in this case would
mean the probability density distribution for initial values of the
3-metric and scalar field in an initial hypersurface, we can extract a lot of
information from Eq. (\ref{hj}) whatever is the quantum potential $Q$,
as we will see. After we get these results, we will return to this
probability issue in the last section.

First we note that, whatever is the form of the quantum potential $Q$,
it must be a scalar density of weight one. This comes from the
Hamilton-Jacobi equation (\ref{hj}). From this equation we can express
$Q$ as
\begin{equation}
Q = -\kappa G_{ijkl}\frac{\delta S}{\delta h_{ij}}
\frac{\delta S}{\delta h_{kl}}
- \frac{1}{2}h^{-1/2} \biggr(\frac{\delta S}{\delta \phi}\biggl)^2 - V .
\end{equation}
As $S$ is an invariant (see Eq. (\ref{smos})), then
$\delta S / \delta h_{ij}$ and $\delta S /\delta \phi$ must be
a second rank tensor density and a scalar density, both of weight one,
respectively. When their products are contracted with $G_{ijkl}$ and
multiplied by $h^{-1/2}$, respectively, they form a scalar density
of weight one. As $V$ is also a scalar density of weight one, then
$Q$ must also be.
Furthermore, $Q$ must depend only on $h_{ij}$ and $\phi$ because it
comes from the wave functional which depends only on these variables.
Of course it can be non-local,
i.e., depending on integrals of the fields over the whole space, but
it cannot depend on the momenta.

A minisuperspace is the set of all spacelike geometries where all but
a set of the $h^{ij}_{(n)}(t)$ and the corresponding $\Pi _{ij}^{(n)}(t)$
are put identically to zero.

Evidently, this procedure violate the uncertainty principle. However,
we expect that the quantization of these minisuperspace models retains
many of the qualitative features of the full quantum theory, which
are easier to study in this simplified model. For more details
on minisuperspace models, see Refs.  \cite{hal1,rya,kucmin}.

In the case of a minisuperspace of homogeneous models, the
supermomentum constraint ${\cal H}^i$ is identically zero, and the shift
function $N_i$ can be set to zero in equation (\ref{hgr}) without loosing
any of the Einstein's equations. The hamiltonian (\ref{hgr}) is
reduced to:
\begin{equation}
\label{homham}
H_{GR} = N(t) {\cal H}(p^{\alpha}(t), q_{\alpha}(t)),
\end{equation}
where $p^{\alpha}(t)$ and $q_{\alpha}(t)$ represent the homogeneous
degrees of freedom coming from $\Pi ^{ij}(x,t)$ and $h_{ij}(x,t)$.
Equations (\ref{hj}-\ref{grf}) become:
\begin{equation}
\label{hoqg}
\frac{1}{2}f_{\alpha\beta}(q_{\mu})\frac{\partial S}{\partial q_{\alpha}}
\frac{\partial S}{\partial q_{\beta}}+ U(q_{\mu}) +
Q(q_{\mu}) = 0,
\end{equation}
\begin{equation}
\label{hqgqp}
Q(q_{\mu}) = -\frac{1}{R} f_{\alpha\beta}\frac{\partial ^2 R}
{\partial q_{\alpha} \partial q_{\beta}},
\end{equation}
\begin{equation}
\label{h}
p^{\alpha} = \frac{\partial S}{\partial q_{\alpha}} =
f^{\alpha\beta}\frac{1}{N}\frac{\partial q_{\beta}}{\partial t} = 0,
\end{equation}
where $f_{\alpha\beta}(q_{\mu})$ and $U(q_{\mu})$ are the minisuperspace
particularizations of $G_{ijkl}$ and $-h^{1/2}R^{(3)}(h_{ij})$, respectively.

Equation (\ref{h}) is invariant under time reparametrization. Hence,
even at the quantum level, different choices of $N(t)$ yield the same
spacetime geometry for a given non-classical solution $q_{\alpha}(x,t)$.

\section{AVOIDANCE OF SINGULARITIES AND QUANTUM ISOTROPIZATION OF THE UNIVERSE}

One of the fluids which may represent the matter content of the
very early Universe is a massless free scalar field, which is
equivalent to stiff matter \cite{zel'dovich} ($p=\rho$, sound
velocity equal to the speed of light). As for this type of matter
content in a universe which is spatially homogeneous and isotropic
(an excellent approximation for the very early Universe),
$\rho\propto a^{-6}(t)$,
where $a(t)$ is the scale factor of the homogeneous and isotropic
hypersurfaces, in the very early Universe, where $a(t)$ approaches zero,
this term dominates over radiation and dust, whose energy densities
depends on $a(t)$ as $a^{-4}(t)$ and $a^{-3}(t)$, respectively.
We will concentrate on this
model now. If this scalar field is not present, radiation will be the dominant
term in the early Universe. This case is studied in detail in
Ref.\cite{bola2}.

Let us take the lagrangian
\begin{equation}
\label{lg1}
{\it L} = \sqrt{-g}\biggr(R - \frac{1}{2}\phi_{,\rho}\phi^{,\rho}\biggl)
\quad,
\end{equation}
where $R$ is the Ricci scalar of the metric $g_{\mu\nu}$ with determinant $g$, and
$\phi$ is the scalar field.
The gravitational part of the minisuperspace model is given by the homogeneous and anisotropic Bianchi I line element
\begin{eqnarray} \label{bia}
ds^{2} &&=-N^{2}(t)dt^{2}+\exp [2\beta _{0}(t)+2\beta _{+}(t)+2\sqrt{3}\beta
_{-}(t)]\;dx^{2}+ \nonumber \\
&&\exp [2\beta _{0}(t)+2\beta _{+}(t)-2\sqrt{3}\beta _{-}(t)]\;dy^{2}+
\exp [2\beta _{0}(t)-4\beta _{+}(t)]\;dz^{2} \;.
\end{eqnarray}
This line element will be isotropic if and only if $\beta _{+}(t)$ and
$\beta _{-}(t)$ are constants \cite{he}.  Other authors have studied Bianchi IX models adopting other interpretations of
quantum cosmology \cite {mos,ams,ber,vil,mon}.

Inserting Equation (\ref{bia}) into the
action $S=\int {{\it L\,}d^{4}x}$,
supposing that the scalar field $\phi $
depends only on time, discarding surface terms, and performing a Legendre
transformation, we obtain the following minisuperspace classical Hamiltonian
\begin{equation} \label{hbiaf}
H=\frac{N}{24\exp {(3\beta _{0})}}(-p_{0}^{2}+p_{+}^{2}+p_{-}^{2}+p_{\phi
}^{2}) \;,
\end{equation}
where $(p_{0},p_{+},p_{-},p_{\phi })$ are canonically conjugate to $(\beta
_{0},\beta _{+},\beta _{-},\phi )$, respectively, and we made the trivial
redefinition $\phi \rightarrow \sqrt{C_{w}/6}\;\;\phi $.

We can write this Hamiltonian in a compact form by defining $y^{\mu} =
(\beta _0, \beta _+, \beta _-, \phi)$ and their canonical momenta $p_{\mu} =
(p_0, p_+, p_-, p_{\phi})$, obtaining
\begin{equation}  \label{ham}
H = \frac{N}{24 \exp{(3y^0)}}\eta ^{\mu\nu}p_{\mu}p_{\nu} \;,
\end{equation}
where $\eta ^{\mu\nu}$ is the Minkowski metric with signature $(-+++)$. The
equations of motion are the constraint equation obtained by varying the
Hamiltonian with respect to the lapse function $N$

\begin{equation}  \label{hbia1f}
{\cal H} \equiv \eta ^{\mu\nu}p_{\mu}p_{\nu} = 0 \;,
\end{equation}
and the Hamilton's equations

\begin{equation}  \label{hbia2f}
\dot{y}^{\mu} = \frac{\partial{\cal H}}{\partial p_{\mu}} = \frac{N}{12 \exp{
(3y_0)}}\eta ^{\mu\nu}p_{\nu} \;,
\end{equation}

\begin{equation}  \label{hbia3f}
\dot{p}_{\mu} = -\frac{\partial{\cal H}}{\partial y^{\mu}} = 0 \;.
\end{equation}
The solution to these equations in the gauge $N=12\exp (3 y_0)$ is

\begin{equation}  \label{solc}
y^{\mu} = \eta ^{\mu\nu}p_{\nu}t + C^{\mu} \;,
\end{equation}
where the momenta $p_{\nu}$ are constants due to the equations of motion and
the $C^{\mu}$ are integration constants. We can see that the only way to
obtain isotropy in these solutions is by making $p_{1}=p_{+}=0$ and $
p_{2}=p_{-}=0$, which yields solutions that are always isotropic, the usual
Friedmann-Robertson-Walker (FRW) solutions with a scalar field. Hence, there
is no anisotropic solution in this model which can classically becomes
isotropic during the course of its evolution. Once anisotropic, always
anisotropic. If we suppress the $\phi$ degree of freedom, the unique
isotropic solution is flat space-time because in this case the constraint (%
\ref{hbia1f}) enforces $p_0 =0$.

To discuss the appearance of singularities, we need the Weyl square tensor $
W^{2}\equiv W^{\alpha \beta \mu \nu }W_{\alpha \beta \mu \nu }$. It reads
\begin{equation} \label{w2}
W^{2}=\frac{1}{432}e^{-12\beta
_{0}}(2p_{0}p_{+}^{3}-6p_{0}p_{-}^{2}p_{+}+p_{-}^{4}+2p_{+}^{2}p_{-}^{2}+
p_{+}^{4}+p_{0}^{2}p_{+}^{2}+p_{0}^{2}p_{-}^{2}) \;.
\end{equation}
Hence, the Weyl square tensor is proportional to $\exp {(-12\beta _{0})}$
because the $p$'s are constants (see Equations (\ref{hbia3f})) and the singularity
is at $t=-\infty $. The classical singularity can be avoided only if we set $
p_{0}=0$. But then, due to Equation (\ref{hbia1f}), we would also have $
p_{i}=0$, which corresponds to the trivial case of flat space-time. Therefore,
the unique classical solution which is non-singular is the trivial flat
space-time solution.

The Dirac quantization procedure yields the Wheeler-DeWitt equation through
the imposition of the condition
\begin{equation}  \label{nense}
\hat{{\cal H}} \Psi = 0 \;,
\end{equation}
on the quantum states, with $\hat{{\cal H}}$ defined as in Equation
(\ref{hbia1f}) (we are assuming the covariant factor ordering) using
the substitutions
\begin{equation}
p_{\mu} \rightarrow - i\frac{\partial }{\partial y^{\mu}} \;.
\end{equation}
Equation (\ref{nense}) reads
\begin{equation}  \label{wdw40}
\eta ^{\mu\nu}\frac{\partial ^2}{\partial y^{\mu} y^{\nu}} \Psi (y^{\mu}) =0
\;.
\end{equation}

For the minisuperspace we are investigating, the guidance relations in the
gauge $N=12\exp (3 y_0)$ are (see Equations (\ref{hbia2f}))

\begin{equation}  \label{gui}
p_{\mu} = \frac{\partial S}{\partial y^{\mu}} = \eta _{\mu\nu}{\dot{y}}
^{\nu} \;,
\end{equation}
where $S$ is the phase of the wave function.

I will
investigate spherical-wave solutions of Equation (\ref{wdw40}). They read

\begin{equation}  \label{psi3}
\Psi_1 = \frac{1}{y}\biggl[ f(y^0 + y) + g(y^0 - y)\biggr] \;,
\end{equation}
where $y\equiv \sqrt{\sum _{i=1}^{3} (y^i)^2}$.

One particular example is the Gaussian superposition of plane wave solutions
of Equation (\ref{wdw40}),
\begin{equation} \label{psi4int}
\Psi _{2}(y^{\mu })=\int {\{F(\vec{k})\exp [i(|\vec{k}|y^{0}+\vec{k}.\vec{y}
)]+G(\vec{k})\exp [i(|\vec{k}|y^{0}-\vec{k}.\vec{y})]\}d^{3}k} \;,
\end{equation}
where $\vec{k}\equiv (k_{1},k_{2},k_{3})$, $\vec{y}\equiv
(y^{1},y^{2},y^{3}) $, $|\vec{k}|\equiv \sqrt{\sum_{i=1}^{3}(k_{i})^{2}}$,
with $F(\vec{k})$ and $G(\vec{k})$ given by
\begin{equation} \label{gauss4}
F(\vec{k})=G(\vec{k})=\exp \biggr[-\frac{(|\vec{k}|-d)^{2}}{\sigma ^{2}}
\biggl] \;.
\end{equation}
After performing the integration in Equation (\ref{psi4int}) using spherical
coordinates we obtain \cite{gra}

\begin{eqnarray}  \label{psi4}
\Psi _{2}(y^{0},y) &=&\frac{i\pi ^{3/2}}{y}\biggl\{[2d\sigma
+i(y^{0}-y)\sigma ^{3}]\exp \biggl[-\frac{(y^{0}-y)^{2}\sigma ^{2}}{4}\biggr]
\exp [id(y^{0}-y)]\nonumber \\
&&\biggl[1+\Phi \biggl(\frac{d}{\sigma }+i(y^{0}-y)
\frac{\sigma }{2}\biggr)\biggr]  \nonumber  \\
&&-[2d\sigma +i(y^{0}+y)\sigma ^{3}]\exp \biggl[-\frac{(y^{0}+y)^{2}\sigma
^{2}}{4}\biggr]\exp [id(y^{0}+y)] \nonumber \\
&& \biggl[1+\Phi \biggl(\frac{d}{\sigma }+
i(y^{0}+y)\frac{\sigma }{2}\biggr)\biggr]\biggr\} \;,
\end{eqnarray}
where $\Phi (x)\equiv (2/\sqrt{{\pi }})\int_{0}^{x}\exp (-t^{2})dt$ is the
probability integral. The wave function $\Psi _{4}$ is a spherical solution
with the form of Equation (\ref{psi3}) with $g=-f$. In order to simplify $\Psi
_{4}$, we will take the limit $\sigma ^{2}>>d$ and $(y^{0} \pm y)\sigma>>1$ in
Equation (\ref{psi4}) yielding \cite{gra}
\begin{equation} \label{fapprox}
f(z)\approx -\frac{16\pi d}{\sigma ^{2}z^{3}} +
            i 2 \pi\biggl(\frac{2}{z^{2}} + \sigma ^{2}\biggr) \;.
\end{equation}

Let us study the spherical wave solutions (\ref{psi3}) of Equation
(\ref{wdw40}). The guidance relations (\ref{gui}) are

\begin{equation}  \label{gui0'}
p_{0} = \partial _0 S = {\rm Im} \biggl(\frac{\partial _0 \Psi _1}{\Psi _1}
\biggr) = -{\dot{y}}^{0} \;,
\end{equation}
\begin{equation}  \label{guii'}
p_{i} = \partial _i S = {\rm Im} \biggl(\frac{\partial _i \Psi _1}{\Psi _1}
\biggr) = {\dot{y}}^{i} \;,
\end{equation}
where $S$ is the phase of the wave function. In terms of $f$ and $g$ the
above equations read

\begin{equation} \label{gui0}
{\dot{y}}^{0}=-{\rm Im}\biggl(\frac{f^{\prime }(y^{0}+y)+g^{\prime }(y^{0}-y)
}{f(y^{0}+y)+g(y^{0}-y)}\biggr) \;,
\end{equation}
\begin{equation} \label{guii}
{\dot{y}}^{i}=\frac{y^{i}}{y}{\rm Im}\biggl(\frac{f^{\prime
}(y^{0}+y)-g^{\prime }(y^{0}-y)}{f(y^{0}+y)+g(y^{0}-y)}\biggr) \;,
\end{equation}
where the prime means derivative with respect to the argument of the
functions $f$ and $g$, and $Im(z)$ is the imaginary part of the complex
number $z$.

From Equations (\ref{guii}) we obtain that

\begin{equation}  \label{yi}
\frac{dy^{i}}{dy^{j}} = \frac{y^i}{y^j} \;,
\end{equation}
which implies that $y^{i}(t)=c_{j}^{i}y^{j}(t)$, with no sum in $j$, where
the $c_{j}^{i}$ are real constants, $c_{j}^{i}=1/c_{i}^{j}$ and $%
c_{1}^{1}=c_{2}^{2}=c_{3}^{3}=1$. Hence, apart some positive multiplicative
constant, knowing about one of the $y^{i}$ means knowing about all $y^{i}$.
Consequently, we can reduce the four equations (\ref{gui0}) and (\ref{guii})
to a planar system by writing $y=C|y^{3}|$, with $C>1$, and working only
with $y^{0}$ and $y^{3}$, say. The planar system now reads

\begin{equation} \label{gui0p}
{\dot{y}}^{0}=-{\rm Im}\biggl(\frac{f^{\prime }(y^{0}+C|y^{3}|)+g^{\prime
}(y^{0}-C|y^{3}|)}{f(y^{0}+C|y^{3}|)+g(y^{0}-C|y^{3}|)}\biggr) \;,
\end{equation}
\begin{equation} \label{guiip}
{\dot{y}}^{3}=\frac{{\rm sign}(y^3)}{C}{\rm Im}\biggl(\frac{f^{\prime
}(y^{0}+C|y^{3}|)-g^{\prime }(y^{0}-C|y^{3}|)}{
f(y^{0}+C|y^{3}|)+g(y^{0}-C|y^{3}|)}\biggr) \;.
\end{equation}
Note that if $f=g$, $y^{3}$ stabilizes at $y^{3}=0$ because ${\dot{y}}^{3}$
as well as all other time derivatives of $y^{3}$ are zero at this line. As $%
y^{i}(t)=c_{j}^{i}y^{j}(t)$, all $y^{i}(t)$ become zero, and the
cosmological model isotropizes forever once $y^{3}$ reaches this line. Of
course one can find solutions where $y^{3}$ never reaches this line, but in
this case there must be some region where ${\dot{y}}^{3}=0$, which implies ${
\dot{y}}^{i}=0$, and this is an isotropic region. Consequently, quantum
anisotropic cosmological models with $f=g$ always have an isotropic phase,
which can become permanent in many cases.

As a concrete example, let us take the Gaussian $\Psi _2$ given in
Equation (\ref{psi4}). It is a spherical wave solution of the Wheeler-DeWitt
equation (\ref{wdw40}) with $f=-g$, and hence it does not necessarily have
isotropic phases as described above for the case $f=g$.
The most interesting case happens when $d$ is negative, as it is shown in
Figure 1 for $d/\sigma^{2}=-10^{-4}$ and $C=2$. Realistic cosmological
models without singularities (in fact, periodic Universes, as one can see
in the closed lines depicted in Figure 1) are obtained,
with expanding phases (increasing $\beta_0$) which
are isotropic [remember that lines parallel to
the $\beta_0$ axes ($\phi\approx$ const.) are also lines with
$\beta_{\pm}\approx$ const. (isotropic phases)],
which can be made arbitrarily large in the region $|\phi |>>|\beta _{0}|$.
Hence, what was classically forbidden
(a nonempty, nonsingular anisotropic model which becomes isotropic in the
course of its evolution)
is possible within the bohmian quantum dynamics described above.

\begin{center}
\includegraphics[
trim=0.035866in 0.035938in 0.000000in 0.036237in,
height=8.2373cm,
width=11.1039cm
]%
{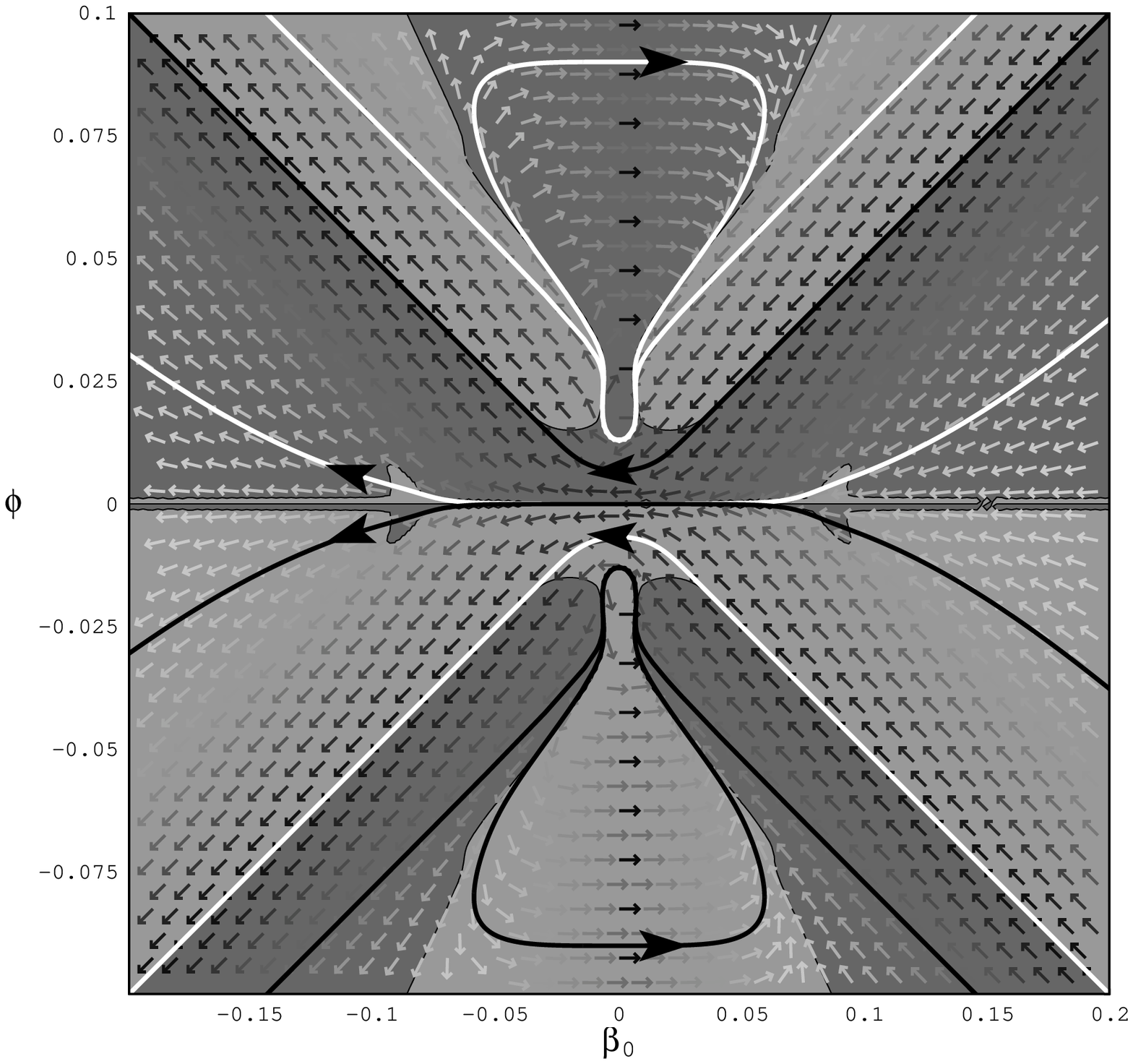}%
\\
\textbf{Figure 1} : Field plot of the system of planar equations (\ref{gui0p}-\ref{guiip}) for
$\sigma =100$, $d=-1$ and $C=2$, which uses the Bohm-de Broglie interpretation
with the wave function $\Psi _2$, Equation (\ref{psi4}). Each arrow of the
vector field is shaded according to its true length, black representing short
vectors and white, long ones. The dark shade of gray shows the regions where
the derivative of the vector field points clockwise (the light shade of gray
means the opposite) and this shading allows to see the regions with arbitrarily
long periods of isotropic evolution. The trajectories are the black or white
curves with direction arrows.
\bigskip
\end{center}

In order to get some analytical insight over Figure 1,
we present the planar system obtained from the guidance relations
corresponding to the wave function (\ref{psi4}) in the approximation
(\ref{fapprox}) :
\begin{equation}
{\dot{\beta}}_{0}=\frac{\frac{2d}{\sigma ^{2}}(3\beta _{0}^{4}+6C^{2}\beta
_{0}^{2}\phi ^{2}-C^{4}\phi ^{4})}{\beta _{0}^{2}(\beta _{0}^{2}-C^{2}\phi
^{2})^{2}+\frac{4d^{2}}{\sigma ^{4}}(3\beta _{0}^{2}+C^{2}\phi ^{2})^{2}} \;,
\end{equation}
\begin{equation}
{\dot{\phi}}=\frac{\frac{16d}{\sigma ^{2}}\beta _{0}^{3}\phi }{\beta
_{0}^{2}(\beta _{0}^{2}-C^{2}\phi ^{2})^{2}+\frac{4d^{2}}{\sigma ^{4}}
(3\beta _{0}^{2}+C^{2}\phi ^{2})^{2}} \;,
\end{equation}
where we have reset $y^{0}=\beta _{0}$ and $y^{3}=\phi$. This approximation
is not reliable in the lines $\beta _{0}=\pm C\phi $. As one can see
immediately from these equations, $\beta _{\pm }={\rm const.}$ whenever
$|\phi |>>|\beta _{0}|$, and the sign of $d$ defines the trajectories
direction.

\section{THE BOHM-DE BROGLIE INTERPRETATION OF FULL QUANTUM SUPERSPACE}

In this section I make general study of the quantum bohmian trajectories
in full superspace.
I will investigate the following important problem. From
the guidance relations (\ref{grh}) and  (\ref{grf}) we obtain
the following first order partial differential equations:
\begin{equation}
\label{hdot}
{\dot{h}}_{ij} =
2NG_{ijkl}\frac{\delta S}{\delta h_{kl}} + D _i N_j + D _j N_i
\end{equation}
and
\begin{equation}
\label{fdot}
\dot{\phi}=Nh^{-1/2}\frac{\delta S}{\delta \phi} + N^i \partial _i \phi .
\end{equation}
The question is, given some initial 3-metric and scalar field,
what kind of structure do we obtain when we integrate this equations
in the parameter $t$? Does this structure form a 4-dimensional
geometry with a scalar field for any choice of the lapse and shift
functions? Note that if the functional $S$ were a solution of the
classical Hamilton-Jacobi equation, which does not contain the quantum
potential term,
then the answer would be in the affirmative because we would be in the
scope of GR. But $S$ is a solution of the {\it modified} Hamilton-Jacobi
equation (\ref{hj}), and we cannot guarantee that this will continue
to be true. We may obtain a complete different structure due to
the quantum effects driven by the quantum potential term
in Eq. (\ref{hj}). To answer this question we will move from
this Hamilton-Jacobi picture of quantum geometrodynamics to a
hamiltonian picture. This is because many strong results concerning
geometrodynamics were obtained in this later picture \cite{kuc1,tei1}.
We will construct a hamiltonian formalism which is consistent with
the guidance relations (\ref{grh}) and (\ref{grf}). It yields the bohmian
trajectories (\ref{hdot}) and (\ref{fdot}) if the guidance relations
are satisfied initially. Once we have this hamiltonian, we can use
well known results in the literature to obtain strong results about
the Bohm-de Broglie view of quantum geometrodynamics.

Examining Eqs. (\ref{smos}) and (\ref{hj}), we can easily guess that
the hamiltonian which generates the bohmian trajectories, once the
guidance relations (\ref{grh}) and (\ref{grf}) are satisfied initially,
should be given by:
\begin{equation}
\label{hq}
H_Q = \int d^3x\biggr[N({\cal H} + Q) + N^i{\cal H}_i\biggl]
\end{equation}
where we define
\begin{equation}
\label{hq0}
{\cal H}_Q \equiv {\cal H} + Q .
\end{equation}
The quantities ${\cal H}$ and ${\cal H}_i$ are the usual
GR super-hamiltonian and
super-momentum constraints given by Eqs. (\ref{h0}) and (\ref{hi}).
In fact, the guidance relations (\ref{grh}) and (\ref{grf}) are consistent
with the constraints ${\cal H}_Q \approx 0$ and ${\cal H}_i \approx 0$
because $S$ satisfies (\ref{smos}) and (\ref{hj}). Futhermore, they are
conserved by the hamiltonian evolution given by (\ref{hq}).
Then we can show that indeed Eqs.(\ref{hdot},\ref{fdot}) can be obtained from $H_Q$
with the guidance relations (\ref{grh}) and (\ref{grf}) viewed as additional
constraints. For details, see Ref.\cite{euc}.

We have a hamiltonian, $H_Q$, which generates the bohmian trajectories
once the guidance relations (\ref{grh}) and (\ref{grf}) are imposed
initially. In the following, we can investigate if the the evolution of the fields
driven by $H_Q$ forms a four-geometry like in classical geometrodynamics.
First we recall a result obtained by Claudio Teitelboim \cite{tei1}.
In this paper, he shows that if the 3-geometries and field configurations
defined on hypersurfaces are evolved by some hamiltonian with the form
\begin{equation}
\label{hg}
\bar{H} = \int d^3x(N\bar{{\cal H}} + N^i\bar{{\cal H}}_i) ,
\end{equation}
and if this evolution can be viewed as the ``motion" of a 3-dimensional
cut in a 4-dimensional spacetime (the 3-geometries can be embedded in
a four-geometry), then the constraints
$\bar{{\cal H}} \approx 0$ and $\bar{{\cal H}}_i
\approx 0$ must satisfy the following algebra

\begin{eqnarray}
\{ \bar{{\cal H}} (x), \bar{{\cal H}} (x')\}&=&-\epsilon[\bar{{\cal
H}}^i(x) {\partial}_i \delta^3(x',x)]
-  \bar{{\cal H}}^i(x') {\partial}_i \delta^3(x,x')
\label{algebra1} \\
\{\bar{{\cal H}}_i(x),\bar{{\cal H}}(x')\} &=& \bar{{\cal H}}(x)
{\partial}_i \delta^3(x,x')
\label{algebra2} \\
\{\bar{{\cal H}}_i(x),\bar{{\cal H}}_j(x')\} &=& \bar{{\cal H}}_i(x)
{\partial}_j \delta^3(x,x')-
\bar{{\cal H}}_j(x') {\partial}_i \delta^3(x,x')
\label{algebra3}
\end{eqnarray}
The constant $\epsilon$ in (\ref{algebra1}) can be $\pm 1$
depending if the four-geometry
in which the 3-geometries are embedded is euclidean
($\epsilon = 1$) or hyperbolic ($\epsilon = -1$).
These are the conditions
for the existence of spacetime.

The above algebra is the same as the algebra (\ref{algebra}) of GR
if we choose $\epsilon = -1$. But the hamiltonian (\ref{hq}) is
different from the hamiltonian of GR only by the presence of
the quantum potential term $Q$ in ${\cal H}_Q$. The Poisson bracket
$\{{\cal H}_i (x),{\cal H}_j (x')\}$ satisfies Eq.
(\ref{algebra3}) because the ${\cal H}_i$ of $H_Q$ defined in Eq.
(\ref{hq}) is the same as in GR. Also
$\{{\cal H}_i (x),{\cal H}_Q (x')\}$ satisfies Eq. (\ref{algebra2})
because ${\cal H}_i$ is the generator of spatial coordinate tranformations,
and as ${\cal H}_Q$ is a scalar density of weight one (remember that
$Q$ must be a scalar density of weight one), then it must satisfies this
Poisson bracket relation with ${\cal H}_i$. What remains to be verified
is if the Poisson bracket
$\{{\cal H}_Q (x),{\cal H}_Q (x')\}$ closes as in Eq. (\ref{algebra1}).
We now recall the result of Ref. \cite{kuc1}. There it is shown that a
general super-hamiltonian $\bar{{\cal H}}$ which satisfies Eq.
(\ref{algebra1}), is a scalar density of weight one, whose geometrical
degrees of freedom are given only by the three-metric $h_{ij}$ and its
canonical momentum, and
contains only even powers and no
non-local term in the momenta (together with the other requirements,
these last two conditions are also satisfied
by ${\cal H}_Q$ because it is quadratic in the momenta and the quantum
potential does not contain any non-local term on the momenta), then
$\bar{{\cal H}}$ must have the following form:

\begin{equation}
\label{h0g}
\bar{{\cal H}} = \kappa G_{ijkl}\Pi ^{ij}\Pi ^{kl} +
\frac{1}{2}h^{-1/2}\pi ^2 _{\phi} + V_G ,
\end{equation}
where

\begin{equation}
\label{vg}
V_G \equiv -\epsilon h^{1/2}\biggl[-{\kappa}^{-1}(R^{(3)} - 2\bar{\Lambda})+
\frac{1}{2}h^{ij}\partial _i \phi\partial _j \phi+\bar{U}(\phi)\biggr] .
\end{equation}
With this result we can now establish three possible scenarios for the
Bohm-de Broglie quantum geometrodynamics, depending on the form of the quantum
potential:
\vspace{1.0cm}

\subsection{Quantum geometrodynamics evolution is consistent and
forms a non degenerate four-geometry}

In this case, the Poisson bracket $\{{\cal H}_Q (x),{\cal H}_Q (x')\}$
must satisfy Eq. (\ref{algebra1}). Then $Q$ must be such that
$V+Q=V_G$ with $V$ given by (\ref{v}) yielding:
\begin{equation}
\label{q4}
Q = -h^{1/2}\biggr[(\epsilon + 1)\biggr(-{\kappa}^{-1} R^{(3)}+
\frac{1}{2}h^{ij}\partial _i \phi\partial _j \phi\biggl)+
\frac{2}{\kappa}(\epsilon\bar{\Lambda} + \Lambda)+
\epsilon\bar{U}(\phi) + U(\phi)\biggl] .
\end{equation}
Then we have two possibilities:

\subsubsection{ The spacetime is hyperbolic ($\epsilon = -1$)}

In this case $Q$ is
\begin{equation}
\label{q4a}
Q = -h^{1/2}\biggr[\frac{2}{\kappa}(-\bar{\Lambda} + \Lambda)
-\bar{U}(\phi) + U(\phi)\biggl] .
\end{equation}
Hence $Q$ is like a classical potential. Its effect is to renormalize the
cosmological constant and the classical scalar field potential, nothing more.
The quantum geometrodynamics is indistinguishable from the classical one.
It is not necessary to require the classical limit $Q=0$ because $V_G=V+Q$
already may describe the classical universe we live in.

\subsubsection{ The spacetime is euclidean ($\epsilon = 1$)}

In this case $Q$ is
\begin{equation}
\label{q4b}
Q = -h^{1/2}\biggr[2\biggr(-{\kappa}^{-1} R^{(3)}+
\frac{1}{2}h^{ij}\partial _i \phi\partial _j \phi\biggl)+
\frac{2}{\kappa}(\bar{\Lambda} + \Lambda)+
\bar{U}(\phi) + U(\phi)\biggl] .
\end{equation}
Now $Q$ not only renormalize the cosmological constant and the
classical scalar field potential but also change the signature of spacetime.
The total potential $V_G=V+Q$ may describe some era of the early universe
when it had euclidean signature,
but not the present era, when it is hyperbolic. The transition between these
two phases must happen in a hypersurface where $Q=0$, which is the classical
limit.

We can conclude from these considerations that if a quantum
spacetime exists with different features from the classical observed one,
then it must be euclidean. In other words, the sole relevant quantum effect
which maintains the non-degenerate nature of the four-geometry of spacetime is its
change of signature to a euclidean one. The other quantum effects are either
irrelevant or break completely the
spacetime structure. This result points in the direction of Ref.
\cite{haw}.
\vspace{1.0cm}

\subsection{Quantum geometrodynamics evolution is consistent but does
not form a non degenerate four-geometry}

In this case, the Poisson bracket $\{{\cal H}_Q (x),{\cal H}_Q (x')\}$
does not satisfy Eq. (\ref{algebra1}) but is weakly zero in some other
way. Let us examine some examples.

\subsubsection{Real solutions of the Wheeler-DeWitt equation}

For real solutions of the Wheeler-DeWitt equation, which is a real
equation, the phase $S$ is null. Then, from Eq. (\ref{hj}), we can
see that $Q=-V$. Hence, the quantum super-hamiltonian
(\ref{hq0}) will contain only the kinetic term, yielding
\begin{equation}
\label{car}
\{{\cal H}_Q (x),{\cal H}_Q (x')\} = 0.
\end{equation}
This is a strong equality. This case is connected with the strong gravity
limit of GR \cite{tei2,hen,san1}. If we take the limit of big gravitational
constant $G$ (or small speed of light $c$, where we arrive at the Carroll group
\cite{poin}), then the potential in the super-hamiltonian constraint of GR
can be neglected and we arrive at a super-hamiltonian containing only
the kinetic term. The Bohm-de Broglie interpretation is telling us
that any real solution of the Wheeler-DeWitt equation yields a quantum
geometrodynamics satisfying precisely this strong gravity limit.
The classical limit $Q=0$ in this case implies also that $V=0$.
It should be interesting to investigate further the structure we obtain here.

\subsubsection{Non-local quantum potentials}

Any non-local quantum potential breaks spacetime.
As an example take a quantum potential
of the form
\begin{equation}
\label{non}
Q=\gamma V ,
\end{equation}
where $\gamma$ is a function of the functional $S$ (here comes the
non-locality). Calculating $\{{\cal H}_Q (x),{\cal H}_Q (x')\}$, we
obtain (see Ref.\cite{euc}),

\begin{eqnarray}
\{ {\cal H}_Q (x), {\cal H}_Q (x')\}&=& (1+\gamma)[{\cal H}^i(x)
{\partial}_i \delta^3(x,x') - {\cal H}^i(x') {\partial}_i \delta^3(x',x)] \nonumber \\
& & - \frac{d \gamma }{d S}V(x')[2{\cal H}_Q (x) - 2 \kappa G_{klij}(x)\Pi^{ij}(x)\Phi^{kl}(x)-
h^{-\frac{1}{2}}\Pi_{\phi}(x)\Phi_{\phi}(x)] \nonumber \\
& & + \frac{d \gamma }{d S}V(x)[2{\cal H}_Q (x')-
2 \kappa G_{klij}(x')\Pi^{ij}(x')\Phi^{kl}(x') - h^{-\frac{1}{2}}\Pi_{\phi}(x')
\Phi_{\phi}(x')] \nonumber \\
& & \approx 0
\end{eqnarray}
The rhs in the last expression is weakly zero because it is a combination
of the constraints and the
guidance relations.
Note that it was very important to use
the guidance relations to close the algebra.
It means that the hamiltonian evolution with the quantum potential
(\ref{non}) is consistent only when restricted to the bohmian trajectories.
For other trajectories, it is inconsitent.

In the examples above, we have explicitly obtained  the
"structure constants" of the algebra that caracterizes
the  ``pre-four-geometry" generated by $H_Q$
i.e., the foam-like structure pointed long time ago in early works
of J. A. Wheeler \cite{whe,whe2}.

Finally, there are no inconsisent bohmian trajectories \cite{sanflu2}.

\section{CONCLUSION}

The Bohm-de Broglie interpretation of canonical
quantum cosmology yields a quantum geometrodynamical picture where
the bohmian quantum evolution
of three-geometries may form, depending on the wave functional, a consistent
non degenerate
four geometry which must be euclidean (but only for a very special local form
of the quantum potential), and a consistent but degenerate four-geometry
indicating the presence of special
vector fields and the breaking of the spacetime structure as a single
entity (in a wider class of possibilities).
Hence, in general, and always when the quantum potential is non-local,
spacetime is broken. The three-geometries evolved under the influence
of a quantum potential do not in general stick together to form a
non degenerate four-geometry, a single spacetime with the causal structure
of relativity. This is not surprising, as it was antecipated long ago
\cite{whe2}. Among the consistent bohmian evolutions, the more general
structures that are formed are degenerate four-geometries with alternative causal structures.
We obtained these results taking a minimally coupled
scalar field as the matter source of gravitation, but
it can be generalized to any matter source with non-derivative
couplings with the metric, like Yang-Mills fields.

As shown in the previous section, a
non degenerate four-geometry can be attained only
if the quantum potential
have the specific form (\ref{q4}).
In this case, the sole relevant quantum effect will be
a change of signature of spacetime, something pointing towards
Hawking's ideas.

In the case of consistent quantum geometrodynamical evolution but with degenerate
four-geometry, we have shown that any real solution of the Wheeler-DeWitt equation
yields a structure which is the idealization of the strong gravity limit
of GR. This type of geometry, which is degenerate, has already been studied
\cite{san1}. Due to the generality of this picture (it is valid for any
real solution of the Wheeler-DeWitt equation, which is a real equation), it
deserves
further attention. It may well be that these degenerate four-metrics were
the correct quantum geometrodynamical description of the young universe.
It would be also
interesting to investigate if these structures have a classical limit
yielding the usual four-geometry of classical cosmology.

As the Bohm-de Broglie interpretation is a more detailed
description of quantum phenomena, in this framework we can investigate
further what kind of structure is formed in quantum
geometrodynamics by using the Poisson
bracket relation (\ref{algebra1}), and the guidance relations
(\ref{hdot}) and (\ref{fdot}). By assuming the existence of 3-geometries,
field configurations, and their momenta, independently on any observations, the
Bohm-de Broglie interpretation allows us to use classical tools, like the
hamiltonian
formalism, to understand the structure of quantum geometry.
The Bohm-de Broglie interpretation yields a lot of information about quantum
geometrodynamics
which cannot be obtained from the many-worlds interpretation
If this
information is useful, I do not know.
However, we cannot answer this question precisely if we
do not investigate further, and the tools are at our disposal.

We would like to remark that all these results were obtained
without assuming any particular factor ordering and regularization
of the Wheeler-DeWitt equation. Also, we did not use any probabilistic
interpretation
of the solutions of the Wheeler-DeWitt equation. Hence, it is a quite general
result. However, we would like to make some comments about the
probability issue in quantum cosmology. The Wheeler-DeWitt equation when applied
to a closed universe does not yield a probabilistic interpretation
for their solutions because of its hyperbolic nature. However, it
has been suggested many times \cite{kow,banks,pad,kie,hal} that at
the semiclassical level we can construct a probability measure with
the solutions of the Wheeler-DeWitt equation. Hence, for interpretations where
probabilities are
essential, the problem of finding a Hilbert space for the solutions
of the Wheeler-DeWitt equation becomes crucial if someone wants to get some
information above the semiclassical level. For instance, at the minisuperspace level,
we fave obtained a lot of information concerning the isotropization of the
Universe and avoidance of singularities in sections III and IV
from quantum cosmological wave equations
which have not a well defined notion of probability.
Of course, probabilities are also useful in the Bohm-de Broglie interpretation.
When we integrate the guidance relations (\ref{hdot}) and (\ref{fdot}), the
initial conditions are arbitrary, and it should be nice to have some
probability distribution on them. However, as we have seen along this contribution,
we can extract a lot of information from the full quantum gravity level
using the Bohm-de Broglie interpretation,
without appealing to any
probabilistic notion. In this interpretation, probabilities are not
essential. Hence, we can take the Wheeler-DeWitt equation as it is, without
imposing any probabilistic interpretation at the most fundamental level,
but still obtaining information using the Bohm-de Broglie interpretation,
and then recover probabilities when we reach the semiclassical level.

It would also be important to investigate the Bohm-de Broglie
interpretation for other
quantum gravitational systems, like black holes. Attempts
in this direction have been made,
but within spherical symmetry in empty space \cite{japa27}, where we
have only a finite number of degrees of freedom.
It should be interesting to investigate more general models. These cases are,
however, qualitatively different from quantum closed cosmological models.
There is no problem in thinking of observers outside an ensemble of
black holes. It is quantum mechanics of an open system, with less conceptual
problems of interpretation.

The conclusions of this paper are of course limited by many strong
assumptions we have tacitly made, as supposing that a continuous three-geometry
exists at the quantum level (quantum effects could also destroy it), or
the validity of quantization of standard GR, forgetting other developments
like string theory. However, even if this approach is not the apropriate one,
it is nice to see how far we can go with the Bohm-de Broglie interpretation,
even in such incomplete stage of canonical quantum gravity.
It seems that the Bohm-de Broglie interpretation may at least be
regarded as a good alternative view to be used in quantum cosmology,
as it will prove harder, or even impossible,
to reach the detailed conclusions of this contribution using other interpretations.
Furthermore, if the finer view of the Bohm-de Broglie interpretation of quantum
cosmology can yield useful information in the form of observational effects,
(as in the possibility raised in Ref.\cite{acc} saying that the present
acceleration of
the Universe may be a quantum cosmological effect obtained from bohmian
quantum dynamics)
then we will have means to
decide between interpretations, something that will be very important not
only for quantum cosmology, but for quantum theory itself.
\vspace{1.0cm}

\section*{ACKNOWLEDGEMENTS}

We would like to thank CNPq of Brazil for financial support and the
group of `Pequeno Semin\'ario' for discussions.

\end{document}